\documentclass[epj]{svjour}
\usepackage{graphics}
\begin{document}
\title{Magneto-Electric response functions for simple atomic systems}
\author{J. Babington \and B. A. van Tiggelen 
}                     
%
%
\institute{Univ. Grenoble 1/CNRS,\\ 
LPMMC UMR 5493, \\
25 rue des Martyrs, Maison des Magist\`{e}res, \\
38042 Grenoble, \\
France. \\
\email{bart.van-tiggelen@grenoble.cnrs.fr, james.babington@grenoble.cnrs.fr. } }
\date{Received: date / Revised version: date}
%
\abstract{
We consider a simple atomic two-body bound state system that is overall charge neutral and placed in a static electric and magnetic field, and calculate the magneto-electric response function as a function of frequency. This is done from first principles using a two-particle Hamiltonian for both an harmonic oscillator and Coulomb binding potential. In the high frequency limit, the response function falls off as $1/\omega^2$ whilst at low frequencies it tends to a constant value.%
} 
\maketitle
\section{Introduction}
\label{intro}

Magneto-electric effects are by now a well established phenomena both theoretically and experimentally. The tendency has been to focus on relatively large molecular systems where DFT calculations~\cite{Rizzo:2003,Rizzo:2009} apply  and experimental values have been measured~\cite{Rikken:2002}. In addition to being associated with important optical phenomena~\cite{Rizzo:2009,Graham:1983}, their existence has played an important role in the Casimir physics~\cite{Feigel:2004zz,Rikken:2006,PhysRevE.76.066605}, in particular if it is possible in certain circumstances to find a contribution from the quantum vacuum to a bodies momentum. For these reasons we consider it an interesting question to ask what are the simplest models that describe possible and display magneto-electric effects.

In this article we calculate the magneto-electric response function for the two simplest bound state systems - the harmonic oscillator and the hydrogen atom. The principal difference between these two atomic systems is that the harmonic oscillator is strongly bound whereas the hydrogen atom with its Coulomb potential is weakly bound. This manifests itself in the accessibility of different energy eigenstates in the perturbation theory

This article is organised as follows. In Section~\ref{sec:gf} we define the atomic system and formulate the response for an arbitrary binding potential. As a tractable example that has a closed form, the harmonic oscillator is chosen and its magneto-electric response function calculated explicitly. In Section~\ref{sec:hydrogen} the response function for hydrogen is presented. It is then compared to the appropriate DFT result and its relation to experimental values. Finally, in Section~\ref{sec:summary} we summarise our results and provide some comment on their validity and applicability.

\section{General formulation}
\label{sec:gf}

We will now derive the ME response function for a two-body charge neutral composite system in static external fields $(\mathbf{E}^0,\mathbf{B}^0)$ using the Hamiltonian formalism (the boldface used here is to indicate they are external fields). One can also use the path integral approach and the coupled classical Lorentz force equations to obtain information about the response function. The path integral gives a correct response function at high frequencies but it is not reliable at low frequencies as well as suffering from the wrong analytic structure in the complex frequency plane~\cite{Wen:2004ym}. A consideration of the coupled Lorentz force equations produces similar difficulties and so we use the Hamiltonian method exclusively in this paper.
\begin{figure}
\begin{center}
\resizebox{0.3\textwidth}{!}{%
\includegraphics{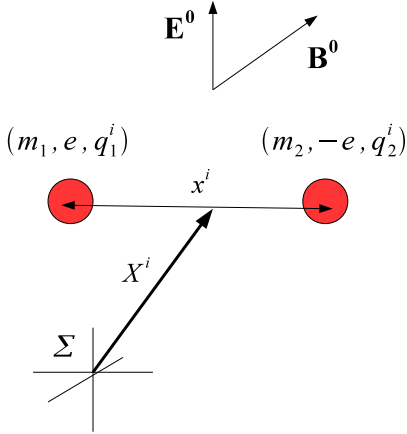} 
}
\end{center}
\caption{The bound state two-body atomic system consisting of two equal but opposite electrical charges, with two different masses $m_1$ and $m_2$ and coordinates $q^i_1$ and $q^i_2$ (with respect to the coordinate system $\Sigma$). To deduce the response function for this system we will change coordinates to the centre of mass coordinate $X^i$ and the separation coordinate $x^i$.} \label{fig:atomicsystem}
\end{figure}

The system we consider is illustrated in Figure~\ref{fig:atomicsystem}. Two equal but opposite electrical charges with coordinates $(q^i_1,q^i_2)$ and masses $(m_1,m_2)$ interacting with a classical (c-number) gauge field $(\phi , A^i)$ that are the combined contribution of the static external fields and the fluctuating source that is used to probe the system. A binding potential $V(q_1 -q_2)$ holds the charges together giving a bound state that is overall charge neutral. The Hamiltonian that describes this system is given by 
\begin{eqnarray}
H&=&\frac{1}{2m_1}(p_1-eA(q_1))^2 +e \phi(q_1) \nonumber \\
&+&\frac{1}{2m_2}(p_2+eA(q_2))^2 -e \phi(q_2)
+ V(q_1 - q_2).\label{eq:hamiltonian1}
\end{eqnarray}
To calculate the response function of this body due to a high frequency electromagnetic field, it is necessary to pass to a new set of coordinates that consists of both the centre of mass coordinate $X^i$ and the separation vector $x^i$. These are shown in Figure~\ref{fig:atomicsystem}. The new variables are then defined by $X := (m_1 q_1 + m_2 q_2)/M $, $x:= (q_1 - q_2)$, $M:= m_1+m_2$, $m:= m_1m_2/M$, and $m_{\Delta}=m_2-m_1$. Correspondingly, we change from the two particles momenta $(p_1,p_2)$ to $(P,p)$ where $P$ is the conjugate momenta of the centre of mass coordinate $X$ and $p$ is likewise the conjugate momenta to the separation vector $x$. It is also necessary to implement this change of coordinates on the gauge field; it can then be expanded about the centre of mass coordinate as
\begin{eqnarray}
A(X+ (m_2/M)x )&=& A(X)+\left(\frac{m_2}{M}\right) x^i\nabla_i A(X) \nonumber \\
&+&\left(\frac{1}{2}\right)\left(\frac{m_2}{M}\right)^2x^ix^j\nabla_i\nabla_j A(X),  \\
A(X- (m_1/M)x )&=& A(X)-\left( \frac{m_1}{M}\right)x^i\nabla_i A(X) \nonumber \\
&+&\left(\frac{1}{2}\right)\left(\frac{m_1}{M}\right)^2x^ix^j\nabla_i \nabla_j A(X) ,
\end{eqnarray}
and similarly for the scalar potential $\phi$. We will work to first order in the spatial derivatives of the gauge potential and thereby neglect the last terms in the above expansion. This corresponds to electric and magnetic fields that can vary in time, but that are spatially constant (so the approach is restricted to wavevectors that are less than the inverse of the size of the atomic system).

Using the Lagrangian as an intermediate step in performing the change of coordinates, we recognize here that $E(t,X)=-\nabla \phi (t,X) -  \partial_{t} A(t,X)$, whilst the derivative of the vector potential once projected with the Levi-Civita tensor will give the magnetic field. The Hamiltonian in the new coordinates after making this expansion reads
\begin{eqnarray}
H &=& \frac{1}{2M}P^2+\frac{1}{2m}p^2+V (x) -ex^i\cdot E_i (t,X) \nonumber \\
&&-e\left(\frac{m_{\Delta}}{M^2}\right) P^i (x\cdot \nabla) A_i(t,X) \nonumber \\
&&-e\left(\frac{m_{\Delta}}{Mm}\right) p^i (x\cdot \nabla )A_i(t,X) \nonumber \\
&&+\frac{e^2}{2m}\left(\frac{m_{\Delta}}{M}\right)^2 x^ix^j\nabla_i A_k(t,X)\nabla_j A_k(t,X)  \nonumber \\
&&+\mathcal{O}(\nabla^2 A).  \label{eq:hamiltonian2} 
\end{eqnarray}
The ME activity results from a source magnetic field inducing electrical polarisation. We can now define the magneto-electric response of the bound state system by promoting all of the canonical degrees of freedom to operators. It is defined by
\begin{equation}
\langle e\delta \hat{x}_i (t) \rangle := \int dt^{\prime}\chi_{ij} ^{EB}(t-t^{\prime})\delta B_j(t^{\prime},X),
\end{equation}
where the lhs is the standard expectation value of the fluctuating electric dipole moment induced on the rhs by an externally applied fluctuating magnetic field (i.e. a test source) that is in general time and space dependent. In terms of correlation functions, it is given by the retarded two-point function
\begin{equation}
\chi_{ij} ^{EB}(t-t^{\prime}) =-i\theta(t-t^{\prime})\langle \Omega \vert [e\delta \hat{x}_i (t),\delta \hat{O}_j(t^{\prime}) ] \vert \Omega \rangle ,
\end{equation}
where the operator $\hat{O}_j(t^{\prime})$ couples to the fluctuating magnetic field $\delta B_j(t^{\prime},X)$, and is to be found from the microscopic theory given by the Hamiltonian Equation~(\ref{eq:hamiltonian2}). The ground state $\vert \Omega \rangle$ that we will use will be specified later in this section. Taking the Fourier transform of this one finds in frequency space
\begin{eqnarray}
\chi_{ij} ^{EB}(\omega) &=&\langle \Omega \vert e\hat{x}_i \frac{1}{\hat{H}-E_0-\hbar \omega}\hat{O}_j  \vert \Omega \rangle \nonumber \\
&&+\langle \Omega \vert e\hat{x}_i \frac{1}{\hat{H}-E_0+\hbar \omega}\hat{O}_j  \vert \Omega \rangle^{\ast},
\end{eqnarray} 
where $E_0$ is the ground state energy of the system. To obtain the detailed form of the response function, we need to specify the form of the operator $\hat{O}_j$. One can see from Equation~(\ref{eq:hamiltonian2}) that the operator has two contributions. One is linear and the other is quadratic in the gauge potential. To calculate the operator we make the specific gauge choice for the gauge potential 
\begin{equation}
A=\frac{1}{2}(\mathbf{B}^0 +\delta B(t) )\wedge X .
\end{equation}
With this choice the Hamiltonian is then a function of only gauge invariant quantities. As a final step to fully specify the Hamiltonian we include the static electric field by substituting $E_i(t,X)=\mathbf{E}_i^0$. The final form of the Hamiltonian that will be used for subsequent calculations is given by
\begin{eqnarray}
H = \frac{1}{2M}\hat{P}^2 +\frac{1}{2m}\hat{p}^2+V (\hat{x}) - e\hat{x}^i\cdot \mathbf{E}^0_i  \nonumber \\
-e\left(\frac{m_{\Delta}}{2M^2}\right) (\hat{x}\wedge \hat{P} )\cdot  (\mathbf{B}^0 +\delta B(t) ) \nonumber \\
-e\left(\frac{m_{\Delta}}{2Mm}\right) \hat{L} \cdot (\mathbf{B}^0 +\delta B(t)) \nonumber \\
+\frac{e^2}{4m}\left(\frac{m_{\Delta}}{M}\right)^2 \hat{x}^i\hat{x}^j\mathbf{B}_m^0 \delta B_n(t)[\delta_{ij}\delta_{mn}-\delta_{im}\delta_{jn}], \label{eq:hamiltonian3} 
\end{eqnarray}
where $L=x\wedge p$ is the orbital angular momentum about the centre of mass origin. One can now just read off the operator that couples to the fluctuating magnetic field
\begin{eqnarray}
\hat{O}_i &=& -\left(\frac{em_{\Delta}}{2mM}\right)\hat{L}_i \nonumber \\
&&+\frac{e^2}{4m}\left(\frac{m_{\Delta}}{M}\right)^2 \hat{x}^a\hat{x}^b [\delta_{ab}\delta_{ij}-\delta_{ai}\delta_{bj}]\textbf{B}^0_j.\label{eq:Moperator}
\end{eqnarray}
There is also a contribution from the term linear in the centre of mass momenta. However, this is zero when evaluated between the ground states (zero momentum plane- wave eigenfunctions). This amounts to a choice of reference frame (the centre of mass frame) and can be used to define the components of the static external electromagnetic fields as well. Indeed, we have not specified so far the nature of the ground state $\vert \Omega \rangle$. For computational convenience we work with a perturbed ground state due to the presence of  static external electric field
\begin{eqnarray}
\vert \Omega \rangle &=& \vert 0 \rangle - e\mathbf{E}_i^0\sum^{\infty}_{n = 1}\frac{1}{E_n-E_0}\vert n \rangle \langle n\vert  \hat{x}^i \vert 0\rangle .
\label{eq:ground state}
\end{eqnarray}
This choice corresponds to the physical situation where we put the two particle system in the static external electric field \emph{first}, let the system settle down and then apply the static external magnetic field. From Equation~(\ref{eq:Moperator}) we see there are two separate contributions to the response function. The first is due to a coupling with the angular momentum operator which we write as $\chi_{ij} ^{EB}(\omega, \hat{L} )$; the second is due to a quadrupole moment coupling which we write as $\chi_{ij} ^{EB}(\omega, \hat{x}^2 )$. Then the full response function is just given by their sum $\chi_{ij} ^{EB}(\omega )=\chi_{ij} ^{EB}(\omega, \hat{L} )+\chi_{ij} ^{EB}(\omega, \hat{x}^2 )$. First consider evaluating the contribution due to the angular momentum operator
\begin{eqnarray}
\chi_{ij} ^{EB}(\omega, \hat{L} ) &=&-\frac{e^2\Delta m}{mM}\langle \Omega \vert  \hat{x}_i \frac{1}{\hat{H}-E_0-\hbar \omega} \hat{L}_j  \vert \Omega \rangle \nonumber \\
&&-\frac{e^2\Delta m}{mM}\langle \Omega \vert  \hat{x}_i \frac{1}{\hat{H}-E_0+\hbar \omega} \hat{L}_j  \vert \Omega \rangle^{\ast} .
\end{eqnarray} 
The next step is to expand the denominators in terms of the static magnetic field. From Equation~(\ref{eq:hamiltonian3}) the perturbation of the Hamiltonian is given by $\delta H =-(em_{\Delta}/mM)\hat{L}_i\textbf{B}_i^0$, therefore we find (keeping only the terms linear in the static magnetic field)
\begin{eqnarray}
\chi_{ij} ^{EB}(\omega , \hat{L}) =-\left(\frac{e^2m_{\Delta} }{mM}\right)\left(\frac{em_{\Delta}}{mM}\right) \nonumber \\
\times\langle \Omega \vert  \hat{x}_i \frac{1}{\hat{H}_0-E_0-\hbar \omega} 
 (\hat{L}_k\textbf{B}_k^0)\frac{1}{\hat{H}_0-E_0-\hbar \omega} \hat{L}_j  \vert \Omega \rangle \nonumber \\
-\left(\frac{e^2m_{\Delta} }{mM}\right)\left(\frac{em_{\Delta}}{mM}\right) \nonumber \\
\times \langle \Omega \vert  \hat{x}_i \frac{1}{\hat{H}_0-E_0+\hbar \omega}
  (\hat{L}_k\textbf{B}_k^0)\frac{1}{\hat{H}_0-E_0+\hbar \omega} \hat{L}_j  \vert \Omega \rangle^{\ast} .
\end{eqnarray} 
Inserting two complete sets of states gives
\begin{eqnarray}
\chi_{ij} ^{EB}(\omega ,\hat{L}) =-\left(\frac{e^3m_{\Delta}^2 }{4m^2M^2}\right)\textbf{B}_k^0\sum_{m,n} \nonumber \\
(\langle \Omega \vert  \hat{x}_i \vert m \rangle 
 \langle m \vert \hat{L}_k \vert n \rangle 
\langle n \vert \hat{L}_j  \vert \Omega \rangle \nonumber \\
\times \frac{1}{E_m-E_0-\hbar \omega} \frac{1}{E_n-E_0-\hbar \omega}  \nonumber \\
+(\langle \Omega \vert  \hat{x}_i \vert m \rangle 
 \langle m \vert \hat{L}_k \vert n \rangle \langle n \vert \hat{L}_j  \vert \Omega \rangle )^{\ast} \nonumber \\
\times \frac{1}{E_m-E_0+\hbar \omega} \frac{1}{E_n-E_0+\hbar \omega} ). \label{eq:MEresponseinter}
\end{eqnarray} 
 
For the quadrupole moment contribution we have
\begin{eqnarray}
\chi_{ij} ^{EB}(\omega, \hat{x}^2) =\left(\frac{e^3m_{\Delta}^2 }{4mM^2}\right) 
 \langle \Omega \vert  \hat{x}_i \left(\frac{1}{\hat{H}-E_0-\hbar \omega}\right) \nonumber \\
\times (\hat{x}^2\mathbf{B}_j^0-\mathbf{B}^0\cdot \hat{x}\hat{x}_j)  \vert \Omega \rangle \nonumber \\
+\left(\frac{e^3m_{\Delta}^2 }{4mM^2} \right)
 \langle \Omega \vert  \hat{x}_i \left(\frac{1}{\hat{H}-E_0+\hbar \omega}\right) \nonumber \\
\times (\hat{x}^2\mathbf{B}_j^0-\mathbf{B}^0\cdot \hat{x}\hat{x}_j)  \vert \Omega \rangle^{\ast}. 
 \label{eq:quad}
 \end{eqnarray}
Inserting a single complete set of states gives 
\begin{eqnarray}
\chi_{ij} ^{EB}(\omega, \hat{x}^2) =\left(\frac{e^3m_{\Delta}^2 }{4mM^2}\right) \sum_n
\left(\frac{1}{E_n-E_0-\hbar \omega}\right)  \nonumber \\
\times \langle \Omega \vert  \hat{x}_i \vert n \rangle  \langle n \vert (\hat{x}^2\mathbf{B}_j^0-\mathbf{B}^0\cdot \hat{x}\hat{x}_j)  \vert \Omega \rangle \nonumber \\
+\left(\frac{e^3m_{\Delta}^2 }{4mM^2} \right) \sum_n \left(\frac{1}{E_n-E_0+\hbar \omega}\right)
  \nonumber \\
\times(\langle \Omega \vert  \hat{x}_i \vert n \rangle \langle n \vert (\hat{x}^2\mathbf{B}_j^0-\mathbf{B}^0\cdot \hat{x}\hat{x}_j)  \vert \Omega \rangle)^{\ast}. 
\label{eq:quad2}
\end{eqnarray}
When the perturbed ground state given by Equation~(\ref{eq:ground state}) is substituted into Equations~(\ref{eq:MEresponseinter}) and~(\ref{eq:quad2}) we arrive at a final expression for the magneto-electric response function. To linear order in the static external fields Equation~(\ref{eq:MEresponseinter}) becomes 

\begin{eqnarray}
\chi_{ij} ^{EB}(\omega ,\hat{L}) =\left(\frac{e^4m_{\Delta}^2 }{4m^2M^2}\right)\textbf{B}_k^0 \mathbf{E}_l^0\sum_{m,n} \sum_{s\neq 0} \nonumber \\
(\langle 0 \vert  \hat{x}_i \vert m \rangle 
 \langle m \vert \hat{L}_k \vert n \rangle 
\langle n \vert \hat{L}_j  \vert s \rangle  \langle s \vert  \hat{x}^l \vert 0 \rangle  \nonumber \\
\times\frac{1}{E_s-E_0} \frac{1}{E_m-E_0-\hbar \omega} \frac{1}{E_n-E_0-\hbar \omega} ) \nonumber \\
+(\langle 0 \vert  \hat{x}_i \vert m \rangle 
 \langle m \vert \hat{L}_k \vert n \rangle \langle n \vert \hat{L}_j   \vert s \rangle  \langle s \vert  \hat{x}^l \vert 0 \rangle  )^{\ast} \nonumber \\
\times\frac{1}{E_s-E_0} \frac{1}{E_m-E_0+\hbar \omega} \frac{1}{E_n-E_0+\hbar \omega} ).  \label{eq:MEresponseL}
\end{eqnarray}
The corresponding form that Equation~(\ref{eq:quad2}) takes is
\begin{eqnarray}
\chi_{ij} ^{EB}(\omega, \hat{x}^2) =-\left(\frac{e^4m_{\Delta}^2 }{4mM^2}\right)\mathbf{B}_k^0\mathbf{E}_l^0 
  \nonumber \\
\times \sum_{n}\sum_{s \neq 0} \left( \frac{1}{E_s-E_0} \frac{1}{E_n-E_0-\hbar \omega} \right) \nonumber \\
\times ((\langle 0 \vert  \hat{x}_i \vert n \rangle  \langle n \vert (\hat{x}^2\delta_{kj} -\hat{x}_k\hat{x}_j)  \vert s \rangle \langle s \vert  \hat{x}^l \vert 0 \rangle \nonumber \\
+\langle s \vert  \hat{x}_i \vert n \rangle  \langle n \vert (\hat{x}^2\delta_{kj} -\hat{x}_k\hat{x}_j)  \vert 0 \rangle \langle 0 \vert  \hat{x}^l \vert s \rangle )\nonumber \\
(\langle 0 \vert  \hat{x}_i \vert n \rangle  \langle n \vert (\hat{x}^2\delta_{kj} -\hat{x}_k\hat{x}_j)  \vert s \rangle  \langle s \vert  \hat{x}^l \vert 0 \rangle \nonumber \\
+\langle s \vert  \hat{x}_i \vert n \rangle  \langle n \vert (\hat{x}^2\delta_{kj} -\hat{x}_k\hat{x}_j)  \vert 0 \rangle \langle 0 \vert  \hat{x}^l \vert s \rangle)^{\ast}). 
\label{eq:MEresponsequadrupole}
\end{eqnarray}

To go further, it is necessary to specify a binding potential so that the energy eigenvalues and eigenfunctions can be deduced.

\subsection{The harmonic oscillator binding potential}

As the simplest example one might consider, the harmonic oscillator with $V(\hat{x})=m\omega^2_0\hat{x}^2/2$. This model is both possible to solve analytically and relevant phenomenologically in displaying the properties associated with real matter. A notable feature here is that since the potential is strongly confining, the perturbed ground state due to the external static electric field has only the first excited state surviving in the summation. Using the operator algebra of the oscillator and the defining relations
\begin{eqnarray}
\hat{x}_i &=& \sqrt{\frac{\hbar}{2m\omega_0}}(\hat{a}^{\dagger}_i+\hat{a}_i) \\
\hat{L}_i &=& -i\hbar/2 \epsilon_{ijk}(\hat{a}_j^{\dagger}\hat{a}_k-\hat{a}_j\hat{a}^{\dagger}_k) \\
&=&-i\hbar \epsilon_{ijk}\hat{a}_j^{\dagger}\hat{a}_k,
\end{eqnarray}
the matrix elements can be evaluated explicitly in Equation~(\ref{eq:MEresponseL})
 \begin{eqnarray}
\chi_{ij} ^{EB}(\omega , \hat{L}) &=&\left(\frac{e^3m_{\Delta}^2 }{m^2M^2}\right)\left(\frac{\hbar^2e}{2m\omega_0^2}\right)\textbf{B}_k^0 
\mathbf{E}_l^0\epsilon_{kib}\epsilon_{jbl} \nonumber \\
 &&\times (\frac{1}{E_1-E_0-\hbar \omega} \frac{1}{E_1-E_0-\hbar \omega}  \nonumber \\
 &&+\frac{1}{E_1-E_0+\hbar \omega} \frac{1}{E_1-E_0+\hbar \omega} ). \nonumber \\
&=&\left(\frac{e^4m_{\Delta}^2 }{\omega^2_0m^3M^2}\right)(
\mathbf{E}_i^0\textbf{B}_j^0 -(\textbf{E}^0 \cdot \mathbf{B}^0)\delta_{ij}) \nonumber \\
&&\times \frac{\omega^2+\omega_0^2}{(\omega_0^2- \omega^2)^2}. \label{eq:MEresponseLHO}
 \end{eqnarray}
In an analogous use of the operator algebra the quadrupole contribution Equation~(\ref{eq:MEresponsequadrupole}) can be similarly evaluated 
\begin{eqnarray}
\chi_{ij} ^{EB}(\omega, \hat{x}^2) &=& -\left(\frac{e^4m_{\Delta}^2 }{4\omega^2_0m^3M^2}\right)
 \left(\frac{1}{\omega_0^2- \omega^2} \right) \nonumber \\
 &&\times (4\mathbf{E}_i^0\textbf{B}_j^0- \mathbf{E}_j^0\textbf{B}_i^0-(\mathbf{E}^0 \cdot\textbf{B}^0)\delta_{ij}). 
 \label{eq:MEresponsequadHO}
 \end{eqnarray}
Note here that both contributions have an anti-symmetric term upon writing the tensor structure in the external fields as a sum of symmetric and antisymmetric pieces, which will have a relevance to our later discussion. Both the angular momentum and quadrupolar contribution are of the same order as can be seen from their multiplicative coefficients. The full response function then takes the final form
\begin{eqnarray}
\chi_{ij} ^{EB}(\omega)=\chi_{ij} ^{EB}(\omega, \hat{L})+ \chi_{ij} ^{EB}(\omega, \hat{x}^2) \nonumber \\
 = -\frac{e^4m_{\Delta}^2 }{\omega^2_0m^3M^2}
[  - \frac{\omega^2+\omega_0^2}{(\omega_0^2- \omega^2)^2}(\mathbf{E}_i^0\textbf{B}_j^0 -(\textbf{E}^0 \cdot \mathbf{B}^0)\delta_{ij})  \nonumber \\
+ \frac{1}{\omega_0^2- \omega^2} (\mathbf{E}_i^0\textbf{B}_j^0- \frac{1}{4} \mathbf{E}_j^0\textbf{B}_i^0-\frac{1}{4}(\mathbf{E}^0 \cdot\textbf{B}^0)\delta_{ij}) ]. \nonumber \\
\label{eq:MEresponseFinal}
\end{eqnarray}

\section{The non-relativistic hydrogen atom}
\label{sec:hydrogen}

\begin{figure}
\begin{center}
\resizebox{0.5\textwidth}{!}{%
\includegraphics{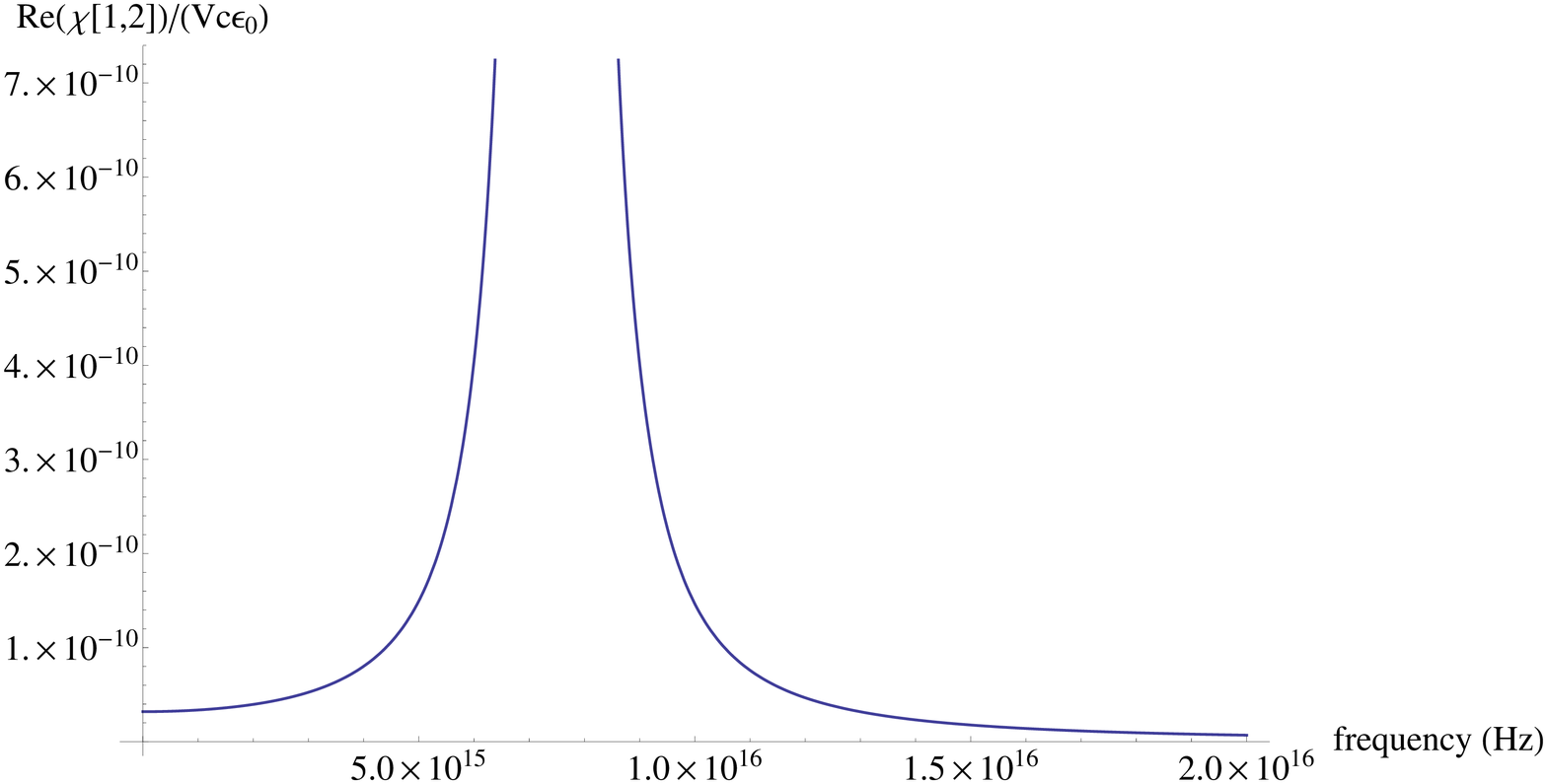} 
}
\end{center}
\caption{The real part of the (dimensionless) off-diagonal susceptibility (the bi-anisotropic component $\chi_{12}$ with $V=(4/3)\pi a_0^3$ the atomic volume in the ground state) of a hydrogen atom with parameters $\mathbf{E^0}=(10^5 Vm^{-1},0,0)$, $\mathbf{B^0}=(0,10T,0)$, $\omega_0=10^{16}Hz$ and $\Gamma = 10^8Hz$. This plot shows the zero and high frequency limits.} \label{fig:plot1}
\end{figure}

\begin{figure}
\begin{center}
\resizebox{0.5\textwidth}{!}{%
\includegraphics{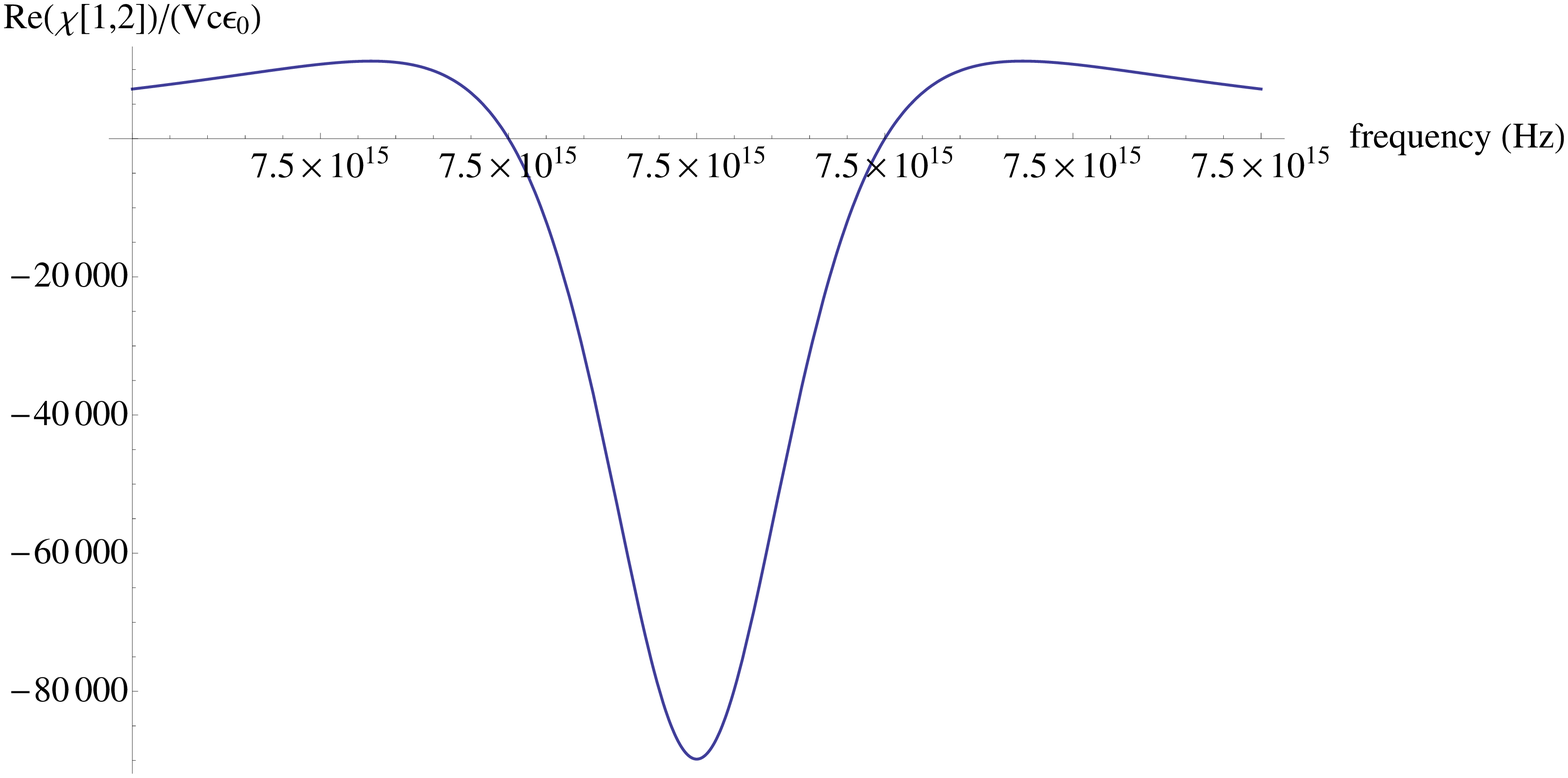} 
}
\end{center}
\caption{The real part of the (dimensionless) off-diagonal susceptibility (the bi-anisotropic component $\chi_{12}$ with $V=(4/3)\pi a_0^3$ the atomic volume in the ground state) of a hydrogen atom with parameters $\mathbf{E^0}=(10^5 Vm^{-1},0,0)$, $\mathbf{B^0}=(0,10T,0)$, $\omega_0=10^{16}Hz$ and $\Gamma = 10^8Hz$. This plot shows the resonant structure.} \label{fig:plot2}
\end{figure}

\begin{figure}
\begin{center}
\resizebox{0.5\textwidth}{!}{%
\includegraphics{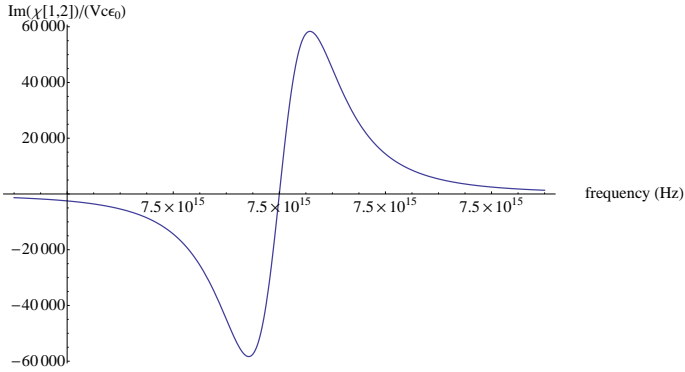} 
}
\end{center}
\caption{The imaginary part of the (dimensionless) off-diagonal susceptibility (the bi-anisotropic component $\chi_{12}$ with $V=(4/3)\pi a_0^3$ the atomic volume in the ground state) of a hydrogen atom with parameters $\mathbf{E^0}=(10^5 Vm^{-1},0,0)$, $\mathbf{B^0}=(0,10T,0)$, $\omega_0=10^{16}Hz$ and $\Gamma = 10^8Hz$. This plot shows the resonant structure.} \label{fig:plot3}
\end{figure}

We now consider the second simplest system, namely the hydrogen atom. This is a weakly bound system with a Coulomb potential given by $V(r)=-e^2/(4\pi \epsilon_0 r)$ in spherical polar coordinates. Since the proton is very much more massive than the electron, we can make the replacement $m_{\Delta}/M \rightarrow 1$, leaving just the reduced mass in all expressions (which is just the electron mass). Equation~(\ref{eq:MEresponseinter}) can be evaluated now using the hydrogenic eigenstates $\vert n, L, m \rangle $ and the energy spectrum $E_n=E_1/n^2$. One finds
\begin{eqnarray}
\chi_{ij} ^{EB}(\omega , \hat{L} ) =-\left(\frac{e^3}{m^2}\right)\textbf{B}_k^0\prod^{2}_{a=1}\sum^{\infty}_{n_a=1}\sum^{n_a-1}_{L_a=0}
\sum^{L_a}_{m_a = -L_a} \nonumber \\
\left( \frac{1}{E_{n_1}-E_1-\hbar \omega} \frac{1}{E_{n_2}-E_1-\hbar \omega} \right) \nonumber \\
\times (\langle \Omega \vert  \hat{x}_i \vert n_1,L_1,m_1 \rangle \langle n_1,L_1,m_1 \vert \hat{L}_k \vert n_2,L_2,m_2 \rangle 
  \nonumber \\
 \times \langle n_2,L_2,m_2 \vert \hat{L}_j  \vert \Omega \rangle \nonumber \\
+\left(\frac{1}{E_{n_1}-E_1+\hbar \omega}  \frac{1}{E_{n_2}-E_1+\hbar \omega} \right) \nonumber \\
(\langle \Omega \vert  \hat{x}_i \vert n_1,L_1,m_1 \rangle \langle n_1,L_1,m_1 \vert \hat{L}_k \vert n_2,L_2,m_2 \rangle  \nonumber \\
 \times \langle n_2,L_2,m_2 \vert \hat{L}_j  \vert \Omega \rangle)^{\ast}. \label{eq:MEresponseLH}
\end{eqnarray}
For the quadrupole contribution Equation~(\ref{eq:quad2}) becomes
\begin{eqnarray}
\chi_{ij} ^{EB}(\omega , \hat{x}^2 ) =+\left(\frac{e^3}{4m^2}\right)\textbf{B}_k^0\sum_{n_1,L_1,m_1}( \langle \Omega \vert  \hat{x}_i \vert n_1,L_1,m_1 \rangle\nonumber \\
\times  \langle n_1,L_1,m_1 \vert \hat{x}^2\delta_{kj}-\hat{x}_k\hat{x}_j \vert \Omega \rangle \frac{1}{E_{n_1}-E_1-\hbar \omega} 
  \nonumber \\
+(\langle \Omega \vert  \hat{x}_i \vert n_1,L_1,m_1 \rangle\nonumber \\
\times  \langle n_1,L_1,m_1 \vert \hat{x}^2\delta_{kj}-\hat{x}_k\hat{x}_j \vert \Omega  \rangle )^{\ast}\frac{1}{E_{n_1}-E_1+\hbar \omega}  . \nonumber \\ \label{eq:MEresponseHQ}
\end{eqnarray}
Because the electron is weakly bound the perturbed ground state due to the external static electric field requires a full summation over the principal quantum number
\begin{eqnarray}
\vert \Omega \rangle &=&  \vert 1,0,0 \rangle - e\mathbf{E}_l^0\sum^{\infty}_{n = 2}\sum^{1}_{m=-1}\frac{1}{E_n-E_1}\vert n, 1, m \rangle \nonumber \\
&&\times \langle n, 1,m \vert \hat{x}_l \vert 1,0,0\rangle ,\label{eq:groundstateh}
\end{eqnarray}
in contrast to the strongly bound harmonic oscillator potential. In Equation~(\ref{eq:MEresponseLH}) the only non-zero elements are when $L_1=L_2=1$ (i.e. just standard selection rules or addition of angular momentum) and thus reduces to
\begin{eqnarray}
\chi_{ij} ^{EB}(\omega , \hat{L}) =\left(\frac{e^4}{m^2}\right)\textbf{B}_k^0\mathbf{E}_l^0 \sum^1_{m_1=-1}\sum^1_{m_2=-1}\sum^{\infty}_{n=2}\sum^{1}_{m=-1} \nonumber \\
\langle 1,0,0 \vert  \hat{x}_i \vert n,1,m_1 \rangle \langle n, 1,m \vert \hat{x}_l \vert 1,0,0\rangle \langle 1,m_1 \vert \hat{L}_k \vert 1,m_2 \rangle 
 \nonumber \\
 \times \langle 1,m_2 \vert \hat{L}_j  \vert 1,m \rangle \frac{1}{(E_{n}-E_1-\hbar \omega)^2} \frac{1}{E_n-E_1} \nonumber \\
+( \langle 1,0,0 \vert  \hat{x}_i \vert n,1,m_1 \rangle\langle n, 1,m \vert \hat{x}_l \vert 1,0,0\rangle 
  \langle 1,m_1 \vert \hat{L}_k \vert 1,m_2 \rangle \nonumber \\
\times \langle 1,m_2 \vert \hat{L}_j  \vert 1,m \rangle)^{\ast}\frac{1}{(E_{n}-E_1+\hbar \omega )^2}\frac{1}{E_n-E_1} . \nonumber \\ \label{eq:MEresponseHL2}
\end{eqnarray}
Turning now to the quadrupole case, Equation~(\ref{eq:MEresponseHQ}) is slightly more complicated as the expectation values of the quadrupole operator requires performing some addition of angular momenta. Indeed, the quadrupole operator can be written as a linear superposition of spherical harmonics $Y_{2,m}$ and $Y_{0,0}$. This implies non-trivial overlaps of matrix elements such that the summation over the $L$ eigenvalue will not completely reduce to a single value as in the previous case but rather one finds
\begin{eqnarray}
\chi_{ij} ^{EB}(\omega , \hat{x}^2 ) &=&-\left(\frac{e^4}{4m^2}\right)\textbf{B}_k^0\textbf{E}_l^0\sum^{\infty}_{n_1=2}\sum^{\infty}_{n = 2}\sum^{3}_{L=1}\sum^{2}_{L_1=1}\sum^{L}_{m=-L} \nonumber \\
&&\times \sum^{L_1}_{m_1=-L_1}\left(\frac{1}{E_n-E_1} \frac{1}{E_{n_1}-E_1-\hbar \omega} \right) \nonumber \\
&&\times( \langle 1,0,0 \vert  \hat{x}_i \vert n_1,L_1,m_1 \rangle \langle n,L, m \vert  \hat{x}_l \vert 1, 0,0 \rangle \nonumber \\
&& \times \langle n_1,L_1,m_1 \vert \hat{x}^2\delta_{kj}-\hat{x}_k\hat{x}_i \vert n, L, m \rangle \nonumber \\
&&\langle 1,0, 0 \vert  \hat{x}_l \vert n,L,m \rangle   \langle n,L, m \vert  \hat{x}_i \vert n_1,L_1,m_1 \rangle \nonumber \\
&&\times  \langle n_1,L_1,m_1 \vert \hat{x}^2\delta_{kj}-\hat{x}_k\hat{x}_j \vert 1, 0, 0 \rangle ) \nonumber \\
&&+\left(\frac{1}{E_n-E_1} \frac{1}{E_{n_1}-E_1+\hbar \omega} \right) \nonumber \\
&&\times( (\langle 1,0,0 \vert  \hat{x}_i \vert n_1,L_1,m_1 \rangle \langle n,L, m \vert  \hat{x}_l \vert 1, 0,0 \rangle \nonumber \\
&& \times \langle n_1,L_1,m_1 \vert \hat{x}^2\delta_{kj}-\hat{x}_k\hat{x}_i \vert n, L, m \rangle )^{\ast}\nonumber \\
&&+(\langle 1,0, 0 \vert  \hat{x}_l \vert n,L,m \rangle   \langle n,L, m \vert  \hat{x}_i \vert n_1,L_1,m_1 \rangle \nonumber \\
&&\times  \langle n_1,L_1,m_1 \vert \hat{x}^2\delta_{kj}-\hat{x}_k\hat{x}_j \vert 1, 0, 0 \rangle)^{\ast} ). \label{eq:MEresponseHQ2}
\end{eqnarray}
Equations~(\ref{eq:MEresponseHL2}) and~(\ref{eq:MEresponseHQ2}) are the main results of this paper. We see that like for the harmonic oscillator, there is a $1/\omega^2$ behaviour at large frequencies. To avoid the singularity when $\hbar \omega = E_n-E_1$,  we supplement a phenomenological line width $\Gamma$ so that any given excited energy level can decay to a lower eigenstate by the replacement $E_n\rightarrow E_n +i \Gamma$. We have evaluated Equations~(\ref{eq:MEresponseHL2}) and~(\ref{eq:MEresponseHQ2}) analytically using \textsf{Mathematica} as an expansion in the principal quantum number. The real part of the susceptibility is plotted in Figures~\ref{fig:plot1} and~\ref{fig:plot2}, as a function of frequency both for the low and high frequency limits, and close to the resonance respectively. To do this we have used numerical data of typical field strengths of $\vert \textbf{B}_k^0\vert=10T$, $\vert \textbf{E}_k^0\vert=10^5Vm^{-1}$, a resonant frequency of $\omega_0=10^{16}Hz$ which corresponds to the energy difference between the first two levels in hydrogen, and a spontaneous decay rate of $\Gamma \sim 10^8 s^{-1}$. In Figure~\ref{fig:plot3} the imaginary part of the susceptibility is plotted close to the resonance. Of course, since the decay rate is very much smaller than the resonant frequencies there is a huge enhancement in the susceptibility close to resonance.

It is worth pointing out here an issue of the validity of our calculation with respect to the photoelectric effect~\cite{Loudon2000}. Given that we are driving an atomic system with an electromagnetic wave at some frequency we may wonder if it is valid at high frequencies. In this regime the associated wave vector is also high and the photon can probe the shorter length scales and transfer more momentum to the electron. However, if one considers the differential cross section for photo-electric effect in hydrogen one knows at high frequency it behaves as $1/\omega^9$. Therefore at high frequency the atom will remain intact and the previous analysis of the magneto-electric response should remain valid.

\subsection{Size of the effect and comparisons}

It is instructive to give a numerical estimate of the size of the response function as compared to the standard electric susceptibility. To do this, consider the harmonic oscillator Equations~(\ref{eq:MEresponseLHO}) and~(\ref{eq:MEresponsequadHO}) at zero frequency. Then up to numerical factors (i.e. simple dimensional analysis) we have
 \begin{eqnarray}
\vert \chi_{ij} ^{EB}(\omega)/(\epsilon_0 c) \vert & \sim &\left(\frac{e^2 }{\epsilon_0 m\omega^2_0}\right)\left(\frac{e^2}{cm^2\omega_0^2}\vert \textbf{B}_k^0\vert \vert
\mathbf{E}_l^0\vert 
 \right) \nonumber \\
 & \sim &\left(\frac{e^2 }{\epsilon_0 m\omega^2_0}\right)\beta
 \label{eq:MEresponse6}
 \end{eqnarray}
The last factor, $\beta$, can be seen too be a dimensionless number, whilst the first is the standard static electric susceptibility. Putting in the number leads to a factor of $\beta\sim 10^{-12}$. This will also be approximately the same for the hydrogen atom if the optical transition frequencies are chosen to coincide.

We can compare this scale to experimental~\cite{Rikken:2002} and DFT  values~\cite{Rizzo:2003} found for the change in refractive $\Delta n$ as compared to the absence of static external electromagnetic fields. For certain large complex molecules values of $\Delta n =(N/V)(\chi/(\epsilon_0 c)) \sim 10^{-11}$ are found experimentally~\cite{Rikken:2002}, where $N/V$ is the number density of the sample. For the helium atom a DFT calculation~\cite{Rizzo:2003} gives a refractive index difference of $\Delta n_{helium} \sim 10^{-17}$ for a sample with number density $N/V \sim 10^{25}m^{-3}$ evaluated at a wavelength of $\lambda = 632.8 nm$. The hydrogen estimate (for the same number density) from our calculations taking into account the number density scaling is $\Delta n_{hydrogen}=(N/V)(\chi/(\epsilon_0 c))\sim (N/V)(e^2/\epsilon_0 m\omega^2_0)\beta\sim 10^{-18}$. It is also worth mentioning that $\vert \chi_{ij} ^{EB}(\omega) \vert $, as for the harmonic oscillator, will scale as size of the system since it is proportional to the standard polarisability. For the helium atom, the static polarisability is $\alpha_{helium}(0) = 0.22\times 10^{-40}$Coulomb meter$^2$/Volt, which when divided by $\epsilon_0$ gives a volume of $16.6 a_0^3$. For the hydrogen atom the corresponding volume is $4a_0^3$ from which we find a scaling factor of approximately four between the hydrogen and helium. The precise form however would have to be fitted empirically with the help of DFT calculations in order to match on to experimental values such as found in~\cite{Rikken:2002}.

\section{Summary}
\label{sec:summary}

We have presented an exact quantum mechanical perturbation theory calculation of the magneto-electric response function for atomic systems with the simplest binding potentials, namely the harmonic oscillator and the Coulomb potential. We have deduced analytic forms for the magneto-electric response tensor as a function of frequency that can be calculated exactly. A common feature is that at high frequency they have $1/\omega^2$ behaviour whilst at low frequency they tend to a constant value. It would be interesting to try and apply the same method exactly to the helium atom.

There are interesting implications of this calculation. Concerning the Feigel effect~\cite{Feigel:2004zz}, where a net momentum density for a medium (Equation (21) in~~\cite{Feigel:2004zz}) with an magneto-electric response function is developed, the result found there is fourth power divergence in frequency, which is then simply cut-off. In  fact it is only the anti-symmetric part of the magneto-electric response tensor that contributes to the momentum. This neglected however the dependence on frequency and was treated as a constant. What we see now is that this divergence will be softened to a quadratic divergence though it will not be simply washed away altogether. This shows that the assumption of a cut-off at high frequencies for ME is not justified and that the divergence has to be resolved by other means, such as done recently in~\cite{Kawka2010}.

\begin{acknowledgement}

\section*{Acknowledgements}

We would like to thank Geert Rikken for useful discussions. This work was supported by the ANR contract PHOTONIMPULS ANR-09-BLAN-0088-01.

\end{acknowledgement}

%
%

\end{document}